\documentclass[10pt]{article}
\usepackage{graphicx}
\usepackage{cite}
\usepackage{color}

\usepackage{amsmath}
\usepackage{amssymb}

\usepackage{bm}

\usepackage{setspace} 
\usepackage[labelfont=bf,labelsep=period,justification=raggedright]{caption}

\makeatletter
\renewcommand{\@biblabel}[1]{\quad#1.}
\makeatother

\date{}

\newcommand{\EQ}{Eq.~}
\newcommand{\EQS}{Eqs.~}
\newcommand{\FIG}{Fig.~}
\newcommand{\FIGS}{Figs.~}

\begin{document}

\begin{flushleft}
{\Large
\textbf{Clustering in large networks does not promote upstream reciprocity}
}
\\
Naoki Masuda$^{1,2,\ast}$, 
\\
\bf{1} Department of Mathematical Informatics,
The University of Tokyo,
Bunkyo, Tokyo, Japan
\\
\bf{2} PRESTO, Japan Science and Technology Agency,
Kawaguchi, Saitama, Japan
\\
$\ast$ E-mail: masuda@mist.i.u-tokyo.ac.jp
\end{flushleft}

\section*{Abstract}

Upstream reciprocity (also called generalized reciprocity) is a putative mechanism for cooperation in social dilemma situations with which players help others when they are helped by somebody else. It is a type of indirect reciprocity. Although upstream reciprocity is often observed in experiments, most theories suggest that it is operative only when players form short cycles such as triangles, implying a small population size, or when it is combined with other mechanisms that promote cooperation on their own. An expectation is that real social networks, which are known to be full of triangles and other short cycles, may accommodate upstream reciprocity. In this study, I extend the upstream reciprocity game proposed for a directed cycle by Boyd and Richerson to the case of general networks. The model is not evolutionary and concerns the conditions under which the unanimity of cooperative players is a Nash equilibrium. I show that an abundance of triangles or other short cycles in a network does little to promote upstream reciprocity. Cooperation is less likely for a larger population size even if triangles are abundant in the network. In addition, in contrast to the results for evolutionary social dilemma games on networks, scale-free networks lead to less cooperation than networks with a homogeneous degree distribution.


\section*{Introduction}\label{sec:introduction}

Several mechanisms govern cooperation among individuals in
social dilemma situations
such as the prisoner's dilemma game. Upstream reciprocity, also called generalized reciprocity,
is one such mechanism in which
players help others when they themselves are helped by other players.
It is a form of indirect reciprocity, in which
individuals are helped by unknown others and vice versa
\cite{Nowak2005nature,Sigmund2010book}.

Cooperation based on upstream
 reciprocity has been observed in various laboratory experiments.
Examples include human subjects in variants
of the trust game, which is a social dilemma game
\cite{Dufwenberg2001HomoOeco,Greiner2005JEconPsy,Stanca2009JEconPsy},
human subjects participating in filling out tedious surveys
\cite{Bartlett2006PsycSci}, and rats
pulling a lever to deliver food to a conspecific
\cite{Rutte2007PlosBiol}.
Even more experimental evidence is available in the field of sociology
in the context of social exchange
\cite{Yamagishi1993SPQ,Molm2007AJS} (also see 
\cite{Malinowski1922book,Ziegler1990chapter}
for classical examples of the Kula ring).

Nevertheless, theory and numerical simulations have revealed that
upstream reciprocity in isolation does not promote cooperation
(but see Barta et al. \cite{Barta2011RoyalB} for 
an exception).
Upstream reciprocity usually supports cooperation only when 
combined with another mechanism that can yield cooperation on its own.
Cooperation appears when the population size is small
\cite{Boyd1989SocN,Pfeiffer2005RoyalB}, upstream reciprocity is combined with
direct reciprocity or spatial reciprocity
\cite{Nowak2007RoyalB},
players move across groups \cite{Hamilton2005RoyalB},
players interact assortatively \cite{Rankin2009Evol},
or players inhabit heterogeneous networks \cite{Iwagami2010JTB}.

\begin{figure}
\begin{center}
\includegraphics[height=3cm]{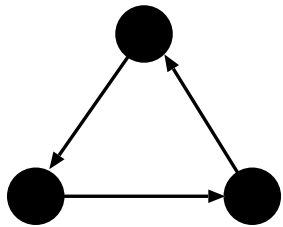}
\end{center}
\caption{{\bf Directed cycle with $N=3$ nodes.}}
\label{fig:directed triangle}
\end{figure}

In their seminal study, Boyd and Richerson analyzed an upstream
reciprocity game on a directed cycle and showed that it yields
cooperation only when the cycle is small
\cite{Boyd1989SocN}. The shortest possible cycle
with indirect reciprocity consists of three players
(\FIG\ref{fig:directed triangle}) because a cycle composed of two
players only involves direct reciprocity.
Cooperation is intuitively less likely for longer cycles
because a player that helps a unique downstream neighbor on the cycle
has to ``trust'' too many intermediary players for their tendency to cooperate before
the player eventually receives help.

Real social networks 
are full of short cycles represented
by triangles, a feature known as transitivity
 \cite{Wasserman94} or clustering \cite{albert02rmp,Newman03siam,Newman10book}.
Therefore, a natural expectation is that larger networks with a high level of clustering
(i.e., many triangles) may facilitate cooperation based on upstream reciprocity
\cite{Yamagishi1993SPQ}.
In the present study, I address this issue theoretically.
I extend the model of Boyd and Richerson \cite{Boyd1989SocN}
to general networks and derive the condition under which the unanimity of
players using upstream reciprocity is resistant to invasion
of defectors.
Then, I apply the
condition to model networks to show that
clustering does little to promote cooperation except in an unrealistic network.
This conclusion holds true for both homogeneous
and heterogeneous networks, where heterogeneity concerns that
in the degree, i.e., the number of neighbors for a player.

My results seem to contradict previous results for spatial
reciprocity in which clustering enhances cooperation in the prisoner's
dilemma game \cite{Nowak1992Nature_spatial} and
those for heterogeneous networks in which heterogeneity
enhances
cooperation in various two-person social dilemma games
\cite{Santos2005PRL,Santos2006PNAS,Santos2006JEB,Duran2005PhysicaD}
and in the upstream reciprocity game \cite{Iwagami2010JTB}.
These previous 
models are evolutionary, however, whereas mine and the original model by
Boyd and Richerson \cite{Boyd1989SocN} are nonevolutionary and based on 
the Nash equilibrium.
I opted to use a nonevolutionary setting in this study because
interpretation of evolutionary games seems elusive
for heterogeneous networks
\cite{Tomassini2007IJMPC,Masuda2007RoyalB} (see Discussion for a more detailed explanation).

\section*{Results}

\subsection*{Preliminary: upstream reciprocity on a directed cycle}\label{sec:Boyd-Richerson}

Boyd and Richerson proposed a model of upstream reciprocity on
the directed cycle \cite{Boyd1989SocN}. By analyzing the
stability of a unanimous population of cooperative players, they showed that cooperation is
unlikely unless the number of players, denoted by $N$, is small.

In their model, the players are involved in a type of donation game.
Each player may donate to a unique downstream neighbor on a directed cycle
at time $t=0$ by paying cost $c (>0)$. The recipient of the donation gains
benefit $b (>c)$. Among the recipients of the donation at $t=0$,
those who comply with upstream reciprocity
donate to a unique downstream
neighbor at $t=1$ by paying cost $c$. Chains of donation
are then carried over to downstream players, who may 
donate to their downstream neighbors at $t=2$. At $t=1$, 
defectors that have received a donation at $t=0$ 
terminate the chain of donation. Such defectors
receive benefit $b$ at $t=0$ and lose nothing at $t=1$.
This procedure is repeated for all players
until all the chains of donation
terminate. If all the players perfectly comply with upstream reciprocity,
the chains never terminate. In contrast, if there is at least one
defector, all the chains terminate in finite time.

As in iterated games \cite{Trivers1971,Axelrod1984book},
$w$ ($0\le w<1$) 
is the probability that the next time step occurs.
We can also interpret $w$ as the probability that players
complying with upstream reciprocity do donate to their downstream neighbors, such that
they erroneously defect with probability $1-w$ in each time step.
Each player's payoff is defined as the discounted 
sum of the payoff over the time horizon. In other words,
the payoff obtained at time $t$ ($\ge 0$)
contributes to the summed payoff
with weight $w^t$.

It may be advantageous for a player not to donate to the downstream neighbor
to gain benefit $b$ without paying cost $c$ over the time course.
However, a player that complies with upstream reciprocity may
enjoy a large summed payoff if chains of donation persist
in the network for a long time.

Each player is assumed to be of either
classical defector (CD; termed unconditional defection
in \cite{Boyd1989SocN}) or
generous cooperator (GC; termed upstream tit-for-tat
in \cite{Boyd1989SocN}).  By
definition, a CD does not donate to the downstream neighbor at $t=0$ and
refuses to relay the chain of donation received from the upstream neighbor
to the downstream neighbor at $t\ge 1$. A GC donates at $t=0$
and donates to the downstream neighbor if the GC received a donation
from the upstream neighbor
in the previous time step.

For this model, Boyd and Richerson
obtained the condition under which
the unanimity of GCs is robust against the
invasion of a CD (i.e., conversion of one GC into CD).
When all players are GC, the summed payoff to one GC
is equal to
\begin{equation}
(b-c)(1+w+w^2+\ldots)=\frac{b-c}{1-w}.
\label{eq:all GC}
\end{equation}
If $N-1$ players are GC and one player is CD, the unique CD's summed payoff
is given by
\begin{equation}b(1+w+w^2+\ldots + w^{N-1})
= \frac{b(1-w^{N-1})}{1-w}.
\label{eq:one CD}
\end{equation}
Therefore, GC is stable against the invasion of CD
if the right-hand side of
\EQ\eqref{eq:all GC} is larger than that of \EQ\eqref{eq:one CD}, that is,
\begin{equation}
w^{N-1}>\frac{c}{b}.
\label{eq:cnd for cycle}
\end{equation}
Equation~\eqref{eq:cnd for cycle} generalizes
the result for direct reciprocity
\cite{Trivers1971,Axelrod1984book},
which corresponds to the case where $N=2$. 
Equation~\eqref{eq:cnd for cycle} also implies that
cooperation is likely if $w$ is large. However, maintaining
cooperation is increasingly difficult as $N$ increases.

\subsection*{Model}

I generalize the Boyd-Richerson model on a directed cycle
to the case of general networks.
Consider a network of $N$ players in which links may be directed or weighted.
I denote the weight of the link from player $i$ to $j$ by $A_{ij}\ge 0$.
I assume that the network is strongly connected, i.e., any player is reacheable from any other player along directed links. Otherwise, chains of donation starting from some playes never return to them because of the purely structural reason. In such a network, it would be more difficult to maintain cooperation than in strongly connected networks. Even for strongly connected networks that might accommodate upstream reciprocity, I will show that cooperation is not likely for realistic network structure.

Assume that all the players are GC and that each GC starts a chain of
donation of unit size at $t=-\infty$. Therefore, the total amount of donation flowing
in the network in each time step is equal to $N$. In the steady state,
the total amount of donation that each player receives from
upstream neighbors is equal to that each player gives to
downstream neighbors in each time step.  I denote the total amount of
donation that reaches and leaves player $i$ by $Nv_i$, where
$\sum_{i=1}^N v_i = 1$. In this situation, the amount of donation that
player $i$ imparts to player $j$ in each time step is equal to
$Nv_iA_{ij}/k_i^{\rm out}$, where $k_i^{\rm out}\equiv \sum_{\ell=1}^N
A_{i\ell}$ is the outdegree of player $i$. Player $i$ receives payoff
$(b-c)Nv_i$ in each time step.

In our previous work
\cite{Iwagami2010JTB}, we assumed that each GC starts a unit flow of 
donation at $t=0$. In the present study, however,
I wait until the flow reaches the steady state
before starting the game at $t=0$.

The definition of CD for general networks is straightforward; a CD
donates to nobody for $t\ge 0$.  I
extend the concept of GC to the case of general networks as follows.
On a directed cycle, a GC quits helping its downstream neighbor
once the GC is not helped by the upstream neighbor \cite{Boyd1989SocN}.
On a general network, the total amount of donation that GC $i$ receives
per unit time in the absence of a CD
is equal to $Nv_i = \sum_{\ell=1}^N Nv_{\ell}A_{\ell
  i}/k_{\ell}^{\rm out}$. If there is a CD, the total amount of donation
that GC $i$ receives may be smaller than the amount that player $i$ would receive
in the absence of a CD.  By
definition, the GC responds to this situation by relaying the
total amount of the incoming donation proportionally to all its downstream
neighbors in accordance with the weights of the links outgoing from player $i$.

As an example, suppose that one upstream neighbor of GC $i$, denoted by
$j$, is CD and all the other $N-1$ players, including player $i$, are
GC.  At $t=0$, the total amount of donation that $i$ receives is equal to
$\sum_{\ell=1,\ell\neq j}^N Nv_{\ell}A_{\ell i}/k_{\ell}^{\rm out}$,
which is smaller than $Nv_i$.  Player $i$ donates
$Nv_i$ in total.  Therefore, player $i$'s payoff at $t=0$ is equal to $b
\sum_{\ell=1,\ell\neq j}^N Nv_{\ell}A_{\ell i}/k_{\ell}^{\rm out} -
cNv_i$. In response to the amount of donation that player $i$ received at $t=0$,
player $i$ adjusts the total amount of donation that it gives the downstream
neighbors from $Nv_i$ to
$\sum_{\ell=1,\ell\neq j}^N Nv_{\ell}A_{\ell i}/k_{\ell}^{\rm out}$
at $t=1$. Therefore,
player $i$ donates
$\sum_{\ell=1,\ell\neq j}^N (Nv_{\ell}A_{\ell
  i}/k_{\ell}^{\rm out})\times (A_{ij^{\prime}}/k_i^{\rm out})$
to its downstream neighbor $j^{\prime}$.
This quantity is smaller than the donation that player $i$
would give player $j^{\prime}$ in the absence of CD $j$, which would be equal to
$Nv_iA_{ij^{\prime}}/k_i^{\rm out}$.

An implicit assumption is that the GC
cannot identify the incoming links along which less donation
is received as compared to the case without a CD.
In other words, even if a GC is defected by the CD in the
upstream, the GC cannot directly retaliate.
Instead, the GC distributes the retaliation equally 
(i.e., proportionally to the weight of the link) to its downstream neighbors.

\subsection*{Stability of upstream reciprocity in networks}\label{sub:analysis}

In this section, I derive the condition under which no player is
motivated to convert from GC to CD when all the players are initially GC.

The steady state $\bm v=(v_1\; \ldots \; v_N)$ is equivalent to
the stationary density of the simple random walk in discrete time.
It is given as the solution of
\begin{equation}
\bm v=\bm v D^{-1}A,
\label{eq:RW}
\end{equation}
where $A=(A_{ij})$ is
the $N$-by-$N$ adjacency matrix, where $A_{ij}$ represents the weight
of the link from $i$ to $j$, and the diagonal
matrix $D$ is defined as
$D={\rm diag}(k_1^{\rm out}, \ldots, k_N^{\rm out})$.
The $(i,j)$ element of $D^{-1}A$ is equal to 
$A_{ij}/k_i^{\rm out}$, that is,
the probability that
a walker at node $i$ transits to node $j$ in one time step.
If the network is undirected, the solution of \EQ\eqref{eq:RW} is given
by $v_i=k_i/\sum_{\ell=1}^N k_{\ell}$, where $k_i=k_i^{\rm
  out}=\sum_{\ell=1}^N A_{i\ell}=\sum_{\ell=1}^N A_{\ell i}$.

The summed payoff to player $i$ is equal to
\begin{equation}
\sum_{t=0}^{\infty} (b-c) w^t Nv_i = \frac{(b-c)Nv_i}{1-w}.
\label{eq:payoff of i when i is GC}
\end{equation}

To examine the Nash stability of the unanimity of GC,
I analyze the situation in which player $i$ is CD and the other
$N-1$ players are GC. At $t=0$, the $N-1$ GCs
pay $cNv_j$ ($j\neq i$), and player $i$ pays nothing.
Therefore, the benefits to the $N$ players, including 
player $i$, at $t=0$ are given in vector form by
\begin{equation}
b N \bm v (I-E_i) D^{-1}A,
\label{eq:benefit t=0}
\end{equation}
where $I$ is the $N$-by-$N$ identity matrix, and
$E_i$ is the $N$-by-$N$ matrix
whose $(i,i)$ element is equal to one and all the
other elements are equal to zero. The benefit to player $j$ ($1\le j\le N$) at $t=0$ is equal to
the $j$th element of the row vector given by \EQ\eqref{eq:benefit t=0}.

At $t=1$, the downstream neighbors of player $i$ donate less because
player $i$ defects at $t=0$.  The amount of donation given to
player $j$, where $j$ is not necessarily a neighbor of $i$, at $t=0$
is equal to the $j$th element of the row vector $N \bm v (I-E_i)
D^{-1}A$. Therefore, the total amount that GC $j (\neq i)$ donates
to its downstream neighbors at $t=1$ is equal to the $j$th element of
$N \bm v (I-E_i) D^{-1}A$. Player $i$, who is CD, does not donate to
others at $t=1$. Therefore, the amount of the donation issued by the
players at $t=1$ is represented in vector form as $N \bm v
(I-E_i)D^{-1}A(I-E_i)$. The discounted benefits that the players
receive at $t=1$ are given in vector form by
\begin{equation}
w bN\bm v \left[(I-E_i)D^{-1}A\right]^2.
\end{equation}
By repeating the same procedure, we can obtain the summed benefits to
the players in vector form as
\begin{equation}
bN\bm v \sum_{t=0}^{\infty} w^t \left[(I-E_i)D^{-1}A\right]^{t+1}
= bN\bm v (I-E_i)D^{-1}A \left[I- w(I-E_i)D^{-1}A\right]^{-1}.
\label{eq:payoff of all when i is CD}
\end{equation}
To derive \EQ\eqref{eq:payoff of all when i is CD},
I used the fact that the spectral radius of $w(I-E_i)D^{-1}A$ is
smaller than unity (that of $D^{-1}A$ is equal to unity).
The $i$th element of \EQ\eqref{eq:payoff of all when i is CD} is
equal to the summed payoff to player $i$ because player $i$ does not pay
the cost to donate at any $t$.

If the $i$th element of
\EQ\eqref{eq:payoff of all when i is CD} is smaller than 
the quantity given by \EQ\eqref{eq:payoff of i when i is GC},
player $i$ is not motivated to turn from GC to CD. Therefore, the 
unanimity of GC is stable if and only if
\begin{equation}
bN\bm v (I-E_i)D^{-1}A \left[I- w(I-E_i)D^{-1}A\right]^{-1}\bigg|_i <
\frac{(b-c)Nv_i}{1-w}\quad (1\le i\le N),
\label{eq:GC stable original}
\end{equation}
where $|_i$ indicates the $i$th element of a vector. By rearranging terms of
\EQ\eqref{eq:GC stable original}, I obtain
\begin{equation}
\bm v\left[I-(I-E_i)D^{-1}A\right]\cdot
\left[I-w(I-E_i)D^{-1}A\right]^{-1}\bigg|_i > \frac{c}{b}v_i\quad
(1\le i\le N).
\label{eq:stability cnd for all-GC}
\end{equation}
Because $\bm v D^{-1}A = \bm v$, \EQ\eqref{eq:stability cnd for all-GC} can be
reduced to
\begin{equation}
\left(\frac{A_{i1}}{k_i}\; \cdots\; \frac{A_{iN}}{k_i}\right)
\left[I-w(I-E_i)D^{-1}A\right]^{-1}\bigg|_i > \frac{c}{b}\quad
(1\le i\le N).
\label{eq:stability cnd for all-GC 2}
\end{equation}

Equation~\eqref{eq:stability cnd for all-GC 2} is never satisfied
when $w=0$ because $A_{ii}=0$. It is
always satisfied when $w\approx 1$ because
the left-hand side of \EQ\eqref{eq:stability cnd for all-GC} tends to $v_i$
as $w\to 1$.

For a directed cycle having $N$ nodes, $\bm v = (1\; \ldots\; 1)/N$,
$k_i^{\rm out} = 1$ ($1\le i\le N$), and $A_{ij}$ is equal to 1 if $(i+1)\mod N=j$
and $0$ otherwise. Owing to the symmetry with
respect to $i$, we only have to consider the 
condition (i.e., \EQ\eqref{eq:GC stable
  original} or
\EQ\eqref{eq:stability cnd for all-GC 2}) for player 1 and obtain the following:
\begin{align}
\bm v (I-E_i)D^{-1}A =& \frac{1}{N}(1\; 0\; 1\; \ldots\; 1),\\
\bm v\left[(I-E_i)D^{-1}A\right]^2 =& \frac{1}{N}(1\; 0\; 0\; 1\; \ldots\; 1),\\
\bm v\left[(I-E_i)D^{-1}A\right]^{N-1} =& \frac{1}{N}(1\; 0\; \ldots\; 0),\\
\bm v\left[(I-E_i)D^{-1}A\right]^N =& (0\; \ldots\; 0).
\end{align}
Therefore, \EQ\eqref{eq:stability cnd for all-GC 2} can be read as
$w^{N-1}>c/b$, which reproduces the result by Boyd and Richerson \cite{Boyd1989SocN}.

\subsection*{Numerical results for various networks}

For general networks, calculating $\left[I-w(I-E_i)D^{-1}A\right]^{-1}$, which is used in
\EQS\eqref{eq:GC stable original} and \eqref{eq:stability cnd for all-GC 2}, is
technically
difficult because this matrix may have nondiagonal Jordan blocks.
Standard formulae for decomposing matrices under
independence of different eigenmodes do not simply apply. The method for
efficiently calculating $\left[I-w(I-E_i)D^{-1}A\right]^{-1}$
is described in the Methods section.

I conducted numerical simulations for different networks to determine
the threshold value of $w$, denoted by $w_{\rm th}$, above which the
unanimity of GC is stable against invasion of CD.  The conclusions
derived from the following numerical simulations are summarized as
follows: (a) abundance of triangles (and other short cycles) hardly
promotes cooperation, and (b) networks with heterogeneous degree
distributions yield less cooperation than those with homogeneous
degree distributions.

\subsubsection*{Network models}

I use five types of undirected networks generated from four network models.
It would be even more 
difficult to obtain cooperation in directed networks because
undirected networks generally allow more direct reciprocity than
directed networks (see Discussion for a more detailed explanation).

The regular random graph (RRG) is defined as
a completely randomly wired network under the restriction that all nodes (i.e.,
players) have
the same degree $k$ \cite{Newman03siam,Newman10book}.
The RRG has low clustering (i.e., low triangle density) 
and is homogeneous in degree
\cite{Watts1998Nature,Newman03siam,Newman10book}.

To construct a network from 
the Watts-Strogatz (WS) model \cite{Watts1998Nature}, nodes are placed
in a circle and connected such that each one is adjacent to the $k/2$
closest nodes on each side on the circle. In this way, each node
has degree $k$. A fraction $p$ of the links is then rewired, and a selected
link preserves one of its end nodes and abandons the other end node. Then, I
randomly select a node from the network as the new destination of the rewired link
such that self-loops and multiple links are avoided.
I use two cases, one in which $p=0$ and the other in which $p$ is small but greater than zero.
In both cases, the network has a high amount of clustering.
When $p=0$, the network is homogeneous in degree
and unrealistic because it has a large average distance between nodes.
When $p$ is positive and appropriately small, the degree is narrowly distributed and the network
has a small average distance \cite{Watts1998Nature}.

As an example of networks with heterogeneous degree
distribution,
I use the Barab\'{a}si-Albert (BA) model.
It has a power-law (scale-free) degree distribution $p(k)\propto k^{-3}$,
a small average distance, and low level of clustering
\cite{Barabasi99sci,albert02rmp}.

To probe the effect of triangles in scale-free networks,
I use a variant of the Klemm-Egu\'{\i}luz (KE)
model \cite{Klemm2002PRE1,Klemm2002PRE2}. For appropriate parameter values,
my variant of the KE model generates scale-free networks
with $p(k)\propto k^{-3}$, small average distances, and a high level of clustering.

\subsubsection*{The effect of clustering}\label{sub:c/b}

For a fixed network and a fixed value of cost-to-benefit ratio $c/b$,
the threshold value of $w$ above which the unanimity of GC is stable against conversion of player $i$ into CD depends on $i$. I denote this value by $w_{\rm th}(i)$.
I determine $w_{\rm th}$
as the largest value of $w_{\rm th}(i)$ ($1\le i\le N$).
This is true because once a certain
player $i$ turns from GC to CD,
other players may be also inclined to
turn to CD. It is straightforward to extend the condition shown in 
\EQ\eqref{eq:GC stable original}
to the case of multiple CD players. For example, 
we can similarly derive the condition under which player $j$ 
turns from GC to CD when player $i$ ($\neq j$) is CD and all the other
$N-2$ players are GC. For example,
on the left-hand side of \EQ\eqref{eq:GC stable original},
we just need to replace $E_i$ with $E_i+E_j$.
I confirmed for all the following numerical results
that once a player turns from GC to CD,
some others are also elicited to turn from GC to CD according to the Nash criterion
and that
such a transition from GC to CD cascades until all players are CD.
In loose terms, this phenomenon is reminiscent of 
models of cascading failure of overloaded networks, 
which mimic, for example, blackouts on power grids
\cite{Motter2002pre}.

\begin{figure}
\begin{center}
\includegraphics[height=6cm]{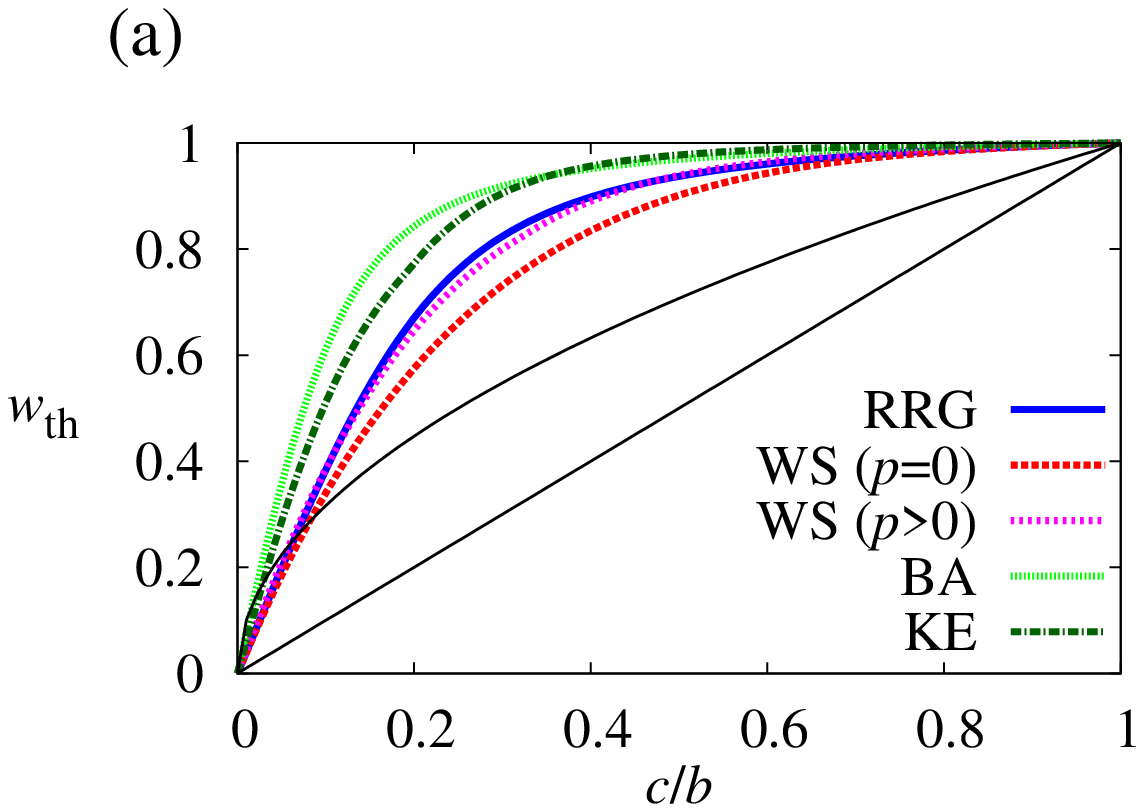}
\includegraphics[height=6cm]{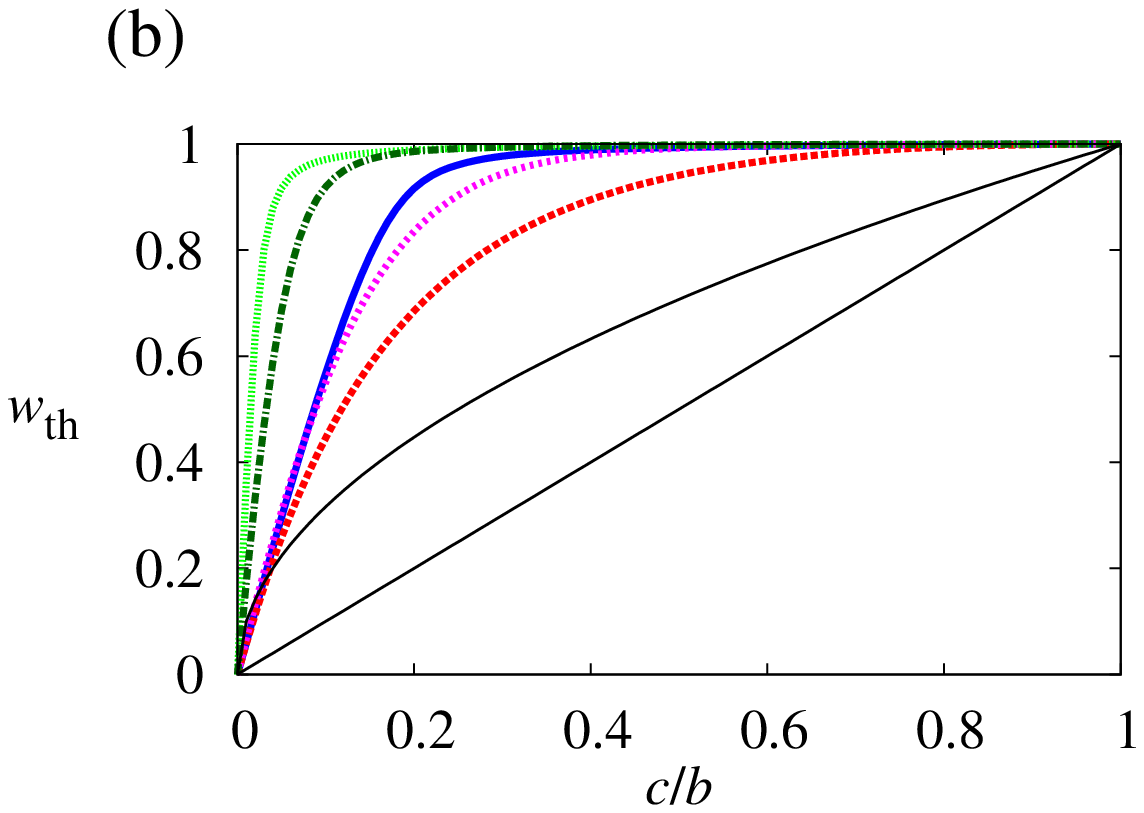}
\end{center}
\caption{
{\bf Relationship between threshold discount factor
($w_{\rm th}$) and cost-to-benefit ratio ($c/b$).}
I use the five types of networks and set
(a) $N=20$, $k=4$, and (b) $N=200$, $k=6$.
The results for direct reciprocity (i.e., $w_{\rm th}=c/b$)
and upstream reciprocity on the directed
triangle (i.e., $w_{\rm th}=\sqrt{c/b}$) are also shown by thin black lines for comparison.
In (a), I set the rewiring
probability for the WS model to $p=0$ and $p=0.1$.
For the BA model, there are initially $m_0=2$ nodes (i.e., dyad),
and the number of links
that each added node has is set to $m=2$. For my variant of 
the KE model, the initial number of nodes and the number of links that each added node has
are set to $m=2$, and
an active node $i$ is deactivated with probability proportional to $(k_i+a)^{-1}$, where
$a=2$. After constructing the network based on the original KE model
\cite{Klemm2002PRE1}, I rewire fraction
$p=0.1$ of randomly selected links to make the average distance small.
In (b), I set $p=0$ and
$p=0.05$ for the WS model, $m_0=m=3$ for the BA model, and $m=a=3$ and
$p=0.05$ for the KE model.}
\label{fig:vary c/b}
\end{figure}

The relationship between $w_{\rm th}$ and
$c/b$ is shown in \FIG\ref{fig:vary c/b}(a)
for the five networks with $N=20$ and mean degree $k=4$. 
The parameter values for the networks are explained in the caption of
\FIG\ref{fig:vary c/b}.
A small $c/b$ value results in a small $w_{\rm th}$ value, indicating
that cooperation is facilitated. This is generally the case
for various mechanisms for cooperation \cite{Nowak2006Science,Sigmund2010book}.

For reference, the results for direct reciprocity ($w_{\rm th}=c/b$) and
upstream reciprocity on the directed triangle (\FIG\ref{fig:directed triangle};
$w_{\rm th}=\sqrt{c/b}$) are also shown in \FIG\ref{fig:vary c/b}(a) by thin black lines.
Except for small $c/b$ values, the five networks with $N=20$ possess higher
$w_{\rm th}$ values as compared to these reference cases.

The two networks generated from the WS model yield smaller values of
$w_{\rm th}$ than those obtained from the RRG, indicating that
the WS model allows more cooperation than the RRG.
Because the degree distributions of these networks are almost the same
and the average distances of the RRG and the WS model with $p=0.1$
do not differ by much \cite{Watts1998Nature},
I ascribe this difference to clustering.
An abundance of triangles and short cycles in networks (i.e., the WS model) enhances cooperation.
However, the difference in $w_{\rm th}$ is not very large. In quantitative terms, clustering
does little to promote cooperation.

The same conclusion is supported for heterogeneous networks
(the BA and KE models). 
Values of
$w_{\rm th}$ for the KE model, which yields a high level of clustering
are smaller than those for the BA model,
which yields a low level of clustering.
However, the $w_{\rm th}$ values for the KE model are considerably larger than those for the
RRG and the WS model, and the
differences between the results for the BA and KE models are small.

To summarize, clustering promotes cooperation but only to a small extent.
To further substantiate this finding, I looked at different cases.
Figure~\ref{fig:vary c/b}(b) compares
$w_{\rm th}$ and $c/b$ values for the networks with $N=200$ and
$k=6$. Figure~\ref{fig:vary N}(a) shows the dependence of
$w_{\rm th}$ on $N$ when $c/b=1/3$.
These cases also suggest that clustering hardly promotes cooperation.

\subsubsection*{Scale-free versus homogeneous networks}

Figure~\ref{fig:vary c/b} indicates that
scale-free networks (i.e., the BA and KE models)
allow less cooperation than networks with
a homogeneous degree distribution (i.e., the RRG and WS model).
This is in contrast with the results for the evolutionary two-person social dilemma games \cite{Santos2005PRL,Santos2006PNAS,Santos2006JEB,Duran2005PhysicaD} and those
for the evolutionary upstream
reciprocity game \cite{Iwagami2010JTB} on heterogeneous networks in
which scale-free networks promote cooperation. The difference stems
from the fact that players in evolutionary games mimic successful
neighbors, whereas in my Nash equilibrium model, players judge
whether GC or CD is more profitable when the other players do not change the strategies
(see Discussion for a more detailed explanation).

\begin{figure}
\begin{center}
\includegraphics[height=6cm]{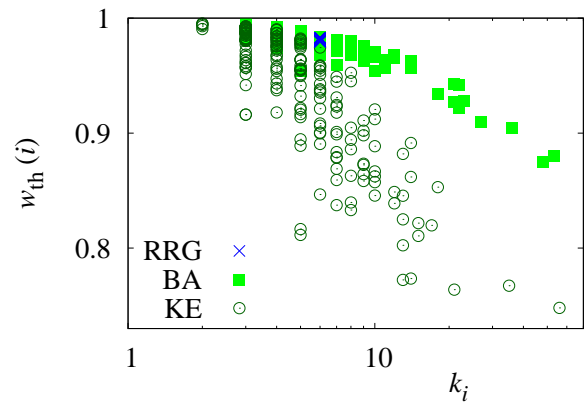}
\end{center}
\caption{
{\bf Relationship between threshold discount factor ($w_{\rm th}(i)$)
and node degree ($k_i$)}. I use the RRG, the BA model, and the KE model with $N=200$ and
$k=6$, and set $c/b=1/3$. The parameter values
for the networks are the same as those used in \FIG\ref{fig:vary c/b}(b).}
\label{fig:degd}
\end{figure}

To probe the reason why cooperation is reduced on scale-free networks,
I examine the dependence of the player-wise 
threshold value, i.e., $w_{\rm th}(i)$ for player $i$, on node degree $k_i$. 
I generate a single network from
each of the RRG, the BA model, and the KE model with $N=200$ using the same
parameter values as those used in \FIG\ref{fig:vary c/b}(b).  For
$c/b=1/3$, the relationship between $w_{\rm th}(i)$ and $k_i$ is
shown in \FIG\ref{fig:degd} for all nodes in the three networks.
$w_{\rm th}(i)$ decreases with $k_i$ in the BA and KE models.
In the RRG, $k_i$ is equal to 6 for all
the nodes, and the value of $w_{\rm th}(i)$ is approximately the same for all the nodes.

$w_{\rm th}(i)$ and $k_i$ are negatively correlated because the amount
of donation flow that a putative CD $i$ stops is
strongly correlated with $v_i$. At $t=0$, it is equal to $v_i$. At
$t\ge 1$, it is generally smaller than $v_i$, but player $i$ having a
large $v_i$ value tends to
receive a large inflow of donation, which player $i$ stops in the next time step.
For undirected networks, $v_i=
k_i/\sum_{\ell=1}^N k_{\ell}\propto k_i$ holds true. 
Players with small degrees are therefore
tempted to convert to CD because the impact of the player's behavior
(i.e., to donate or not to donate) on the entire network is small.
Therefore, a small $k_i$ leads to a large $w_{\rm th}(i)$.
Even for directed networks,
$v_i$ and $k_i^{\rm out}$ are often strongly correlated
\cite{Fortunato2006www,mkk09njp,Ghoshal2011NatComm}. 
Because the minimum degree in a scale-free network is smaller than
that in a homogeneous network if the mean degree of the two networks
is equal, scale-free networks have larger $w_{\rm th}$ as compared to
homogeneous networks.

\subsubsection*{Cooperation in large networks}

A comparison of \FIGS\ref{fig:vary c/b}(a) and \ref{fig:vary
  c/b}(b) suggests that a large $N$ makes cooperation unlikely.  To
examine this point further, I set $c/b=1/3$, generated 100 networks for each
$N$ value and each network type, calculated $w_{\rm th}$, and obtained
the mean and the standard deviation of $w_{\rm th}$.  Because the WS
model with $p=0$ is unique for a given $N$, the mean and standard deviation are not
relevant in this network.

\begin{figure}
\begin{center}
\includegraphics[height=6cm]{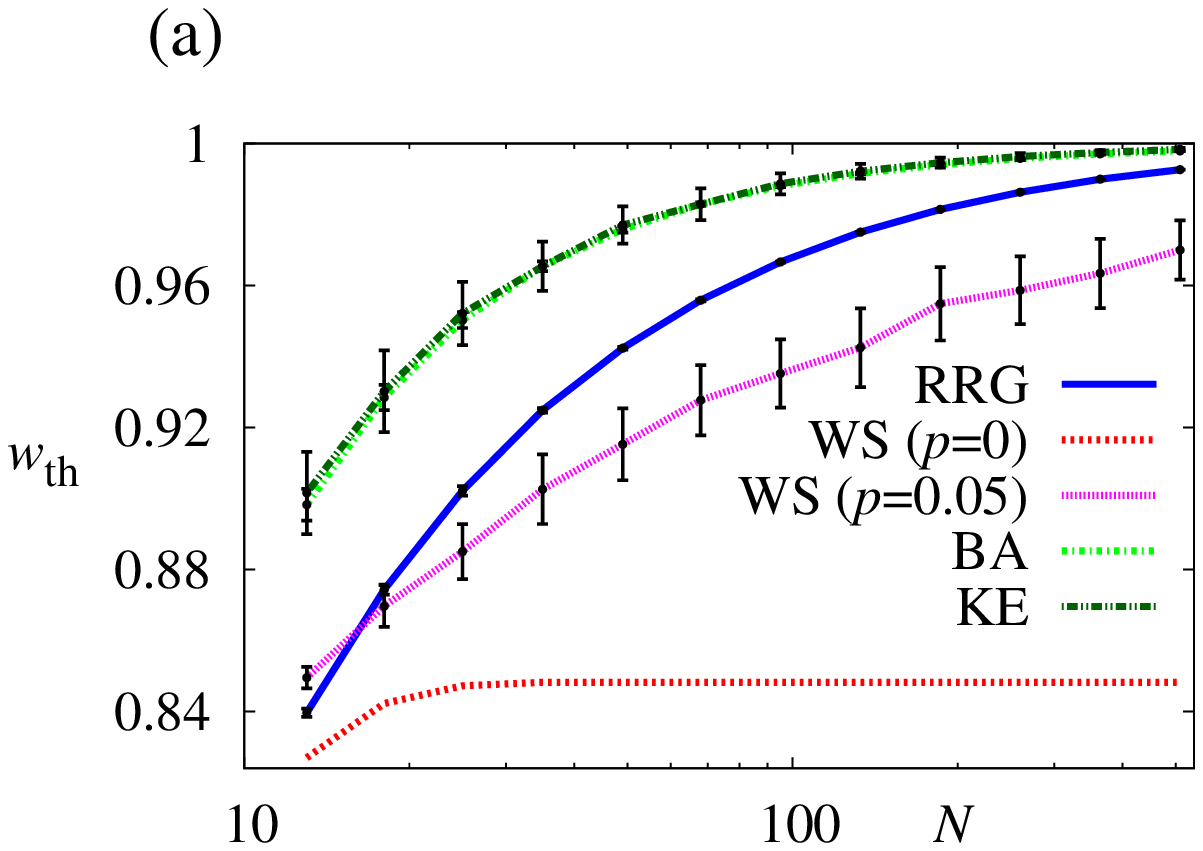}
\includegraphics[height=6cm]{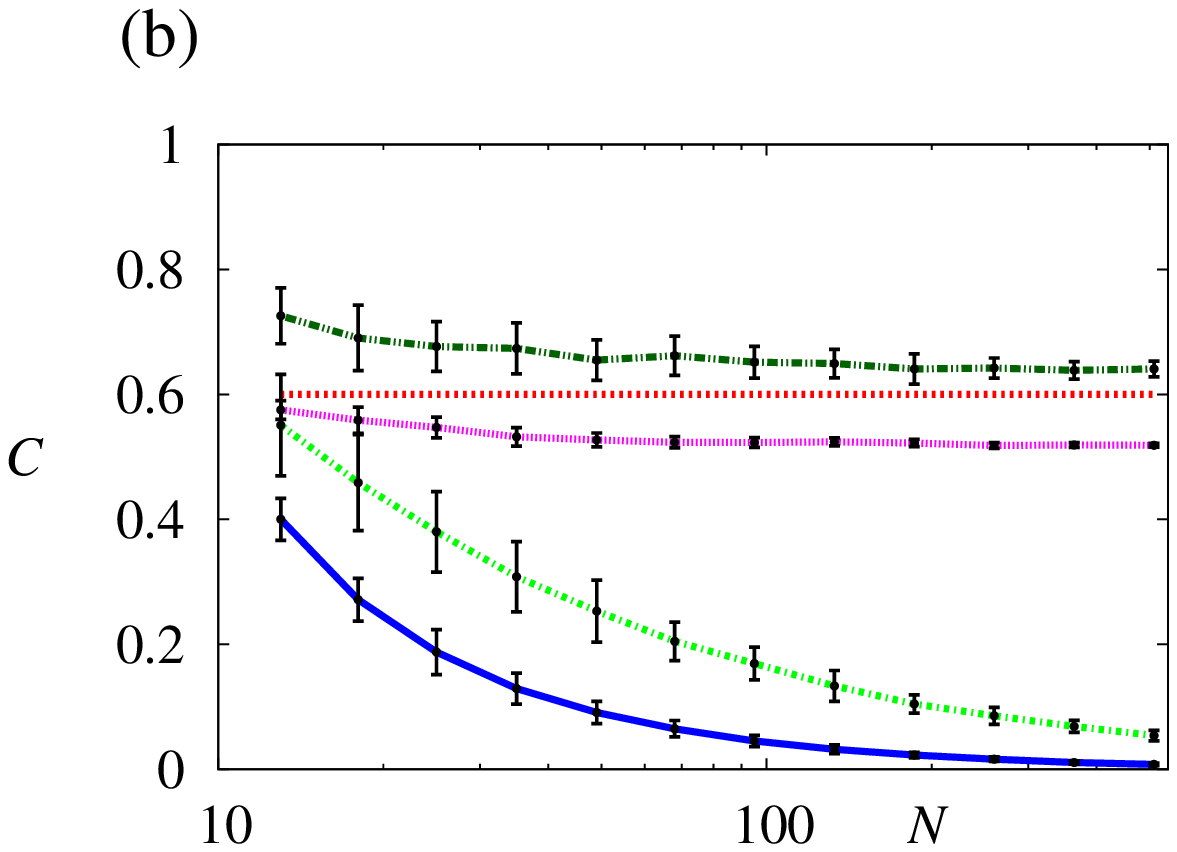}
\end{center}
\caption{
{\bf Effects of network size ($N$).}
(a) Dependence of the threshold discount factor ($w_{\rm th}$) on $N$.
(b) Dependence of the clustering coefficient ($C$) on $N$.
I use the five types of networks and set $c/b=1/3$. The parameter values
for the networks are the same as those used in \FIG\ref{fig:vary c/b}(b).
In (a), the results for the BA and KE models heavily overlap.}
\label{fig:vary N}
\end{figure}

The mean and standard deviation of $w_{\rm th}$ for the five networks
of various sizes are shown in \FIG\ref{fig:vary N}(a).
The results for the BA and KE models heavily overlap.
Cooperation is less likely as $N$ increases in all models, except for the WS model
with $p=0$. This result is consistent with that for a directed cycle
\cite{Boyd1989SocN}.

$w_{\rm th}$ increases with $N$ not
entirely owing to the decreased level of clustering
in the network. To show this, I plot the mean and standard deviation
of the clustering coefficient
$C (\in [0, 1])$, which quantifies the abundance of
triangles in a network \cite{Watts1998Nature},
in \FIG\ref{fig:vary N}(b).
The clustering coefficient is defined as $C\equiv (1/N)\times \sum_{i=1}^N
(\text{number of
  triangles including node } i)/[k_i(k_i-1)/2]$. Figure~\ref{fig:vary
  N}(b) indicates that $C$ decreases with $N$ for the RRG and the BA
model. Therefore, the effect of $N$ and $C$ on $w_{\rm th}$ may be
mixed in these two network models.  However, $C$ stays almost constant
for the WS and KE models. At least for these models,
an increase in $w_{\rm th}$ is considered to originate primarily from an increase in 
$N$, not from changes in the level of clustering.

In \FIG\ref{fig:vary N}(a),
$w_{\rm th}$ seems to approach
unity as $N$ increases except for the WS model with $p=0$.
As previously stated, the WS model with $p=0$ is unrealistic because
it has a large average distance between pairs of nodes
\cite{Watts1998Nature,albert02rmp,Newman03siam,Newman10book}. Therefore, I conclude
that cooperation based on upstream reciprocity is not likely
for homogeneous and heterogeneous networks in general.

\section*{Discussion}

I generalized the upstream reciprocity model proposed for a directed
cycle \cite{Boyd1989SocN} to general networks and reached two primary conclusions.

First, cooperation based on upstream
reciprocity is not likely in general networks regardless of the
abundance of triangles and heterogeneity in the node degree.
Because the networks that I examined are undirected,
some amount of direct reciprocity is relevant;
GC neighbors partially retaliate directly against a CD.
My result that cooperation is unlikely for undirected networks
implies that cooperation would be even more difficult for
directed networks in which direct reciprocity is less available.
In directed networks, direct reciprocity occurs only on reciprocal
links between a pair of players.

Second, I showed
that scale-free network models allow less cooperation
(i.e., large $w_{\rm th}$) as compared to networks
with homogeneous degree distributions.
This result is opposite of those for two-person
social dilemma games
\cite{Santos2005PRL,Santos2006PNAS,Santos2006JEB,Duran2005PhysicaD} and the upstream
reciprocity game \cite{Iwagami2010JTB}.  The difference
stems from the fact that the previous studies assumed evolutionary
games and the present study (and the original model by Boyd and
Richerson \cite{Boyd1989SocN}) is based on nonevolutionary analysis.

I adopted a nonevolutionary setup and examined 
the condition for the Nash equilibrium because the concept of
the evolutionary game on heterogeneous networks seems elusive.
Evolutionary games on heterogeneous networks imply that a player
imitates the strategy of a successful neighbor that is likely to have a
different node degree. However, players with different degrees
are involved in essentially different games
because the number of times that each player plays the game per generation necessarily
depends on the degree.
Therefore, for example, a small-degree player
cannot generally expect a large payoff by
mimicking a successful neighbor with a large degree. In this situation, defining the
game and payoff for players with various degrees is complicated
\cite{Santos2006JEB,Tomassini2007IJMPC,Masuda2007RoyalB}.
Use of the Nash criterion does not incur this type of problem.

The overall conclusions of the present study are negative. To explain
the occurrence of upstream reciprocity in real societies, it may be
advantageous to combine upstream reciprocity
with other non-network mechanisms, such as
the ones mentioned in the Introduction.

\section*{Methods}

\subsection*{Numerical methods for calculating \EQS\eqref{eq:GC stable
original} and \eqref{eq:stability cnd for all-GC 2}}\label{sec:numerical technique}

I determined $w_{\rm th}$ by applying the bisection method
to \EQ\eqref{eq:GC stable original}
or \eqref{eq:stability cnd for all-GC 2}. To
calculate $\left[I-w(I-E_i)D^{-1}A\right]^{-1}$ for different values
of $w$, it is beneficial to use the expansion of $(I-E_i)D^{-1}A$ in terms
of independent modes. This is possible
when the adjacency matrix $A$ for the subnetwork
composed of the GCs is diagonalizable, as shown below.

I assume that
there are $N_{\rm s}\equiv N-N_{\rm d}$ GCs and
$N_{\rm d}$ CDs. 
In the main text,
%
%
I focused on the case $N_{\rm d}=1$.
However, the case $N_{\rm d}\ge 2$ is also relevant because
I verified in the main text that the appearance of a single CD leads to the further
emergence of CDs.
%
%
Without loss of generality, I assume that players 1, 2, \ldots, $N_{\rm s}$ are GC and
players $N_{\rm s}+1$, $N_{\rm s}+2$,
$\ldots$, $N$ are CD, and that
the network is strongly connected. We need to identify
all the (generalized) eigenmodes of $E_{\rm s}D^{-1}A$, where
\begin{equation}
E_{\rm s}\equiv \sum_{i=1}^{N_{\rm s}} E_i.
\label{eq:E_s}
\end{equation}
I first partition $E_{\rm s}$,
$D^{-1}$, and $A$ into two-by-two blocks, each partition corresponding to the
set of GC and that of CD. For a candidate of a left
eigenvector of $E_{\rm s}D^{-1}A$, denoted by $\bm v^{(i)}$,
\begin{align}
\bm v^{(i)} E_{\rm s}D^{-1}A \equiv&
\begin{pmatrix} \bm v_{\rm s}^{(i)} & \bm v_{\rm d}^{(i)} \end{pmatrix}
\begin{pmatrix} I_{N_{\rm s}} & O\\ O & O \end{pmatrix}
\begin{pmatrix} D_{\rm s}^{-1}& O\\ O& D_{\rm d}^{-1}\end{pmatrix}
\begin{pmatrix} A_{\rm ss}& A_{\rm sd}\\ A_{\rm ds}& A_{\rm dd}\end{pmatrix}\notag\\
=& 
\begin{pmatrix} 
\bm v_{\rm s}^{(i)}D_{\rm s}^{-1}A_{\rm ss} & \bm v_{\rm s}^{(i)}D_{\rm s}^{-1}A_{\rm sd}
\end{pmatrix},
\label{eq:try left}
\end{align}
where $I_{N_{\rm s}}$ is the identity matrix of size $N_{\rm s}$;
$D_{\rm s}$ and $D_{\rm d}$ are diagonal matrices whose diagonal entries
are equal to the outdegrees of the GCs and CDs, respectively;
$A_{\rm ss}$ is the $N_{\rm s}$-by-$N_{\rm s}$ matrix corresponding to
the adjacent matrix within the GCs; and 
$A_{\rm sd}$, $A_{\rm ds}$, and $A_{\rm dd}$ are similarly defined
blocks of the original adjacency matrix $A$. 
Note that $A_{\rm ds}$ and $A_{\rm dd}$ are absent on the right-hand
side of
\EQ\eqref{eq:try left} and as such
are not relevant to the following discussion.

First of all, $\bm v^{(i)} = (\bm v_{\rm s}^{(i)}\; \bm v_{\rm
  d}^{(i)}) = \bm e_i^{\top}$ ($N_{\rm s}+1\le i\le N$) is a trivial
zero left eigenvector of $E_{\rm s}D^{-1}A$.  Here, $\top$ denotes
transpose, and $\bm e_i$ is the unit column vector in which the $i$th
element is equal to unity and all the other elements are equal to
zero.

To obtain the other $N_{\rm s}$ generalized eigenmodes of
$E_{\rm s}D^{-1}A$, I consider the case in which $D_{\rm s}^{-1}A_{\rm ss}$ is
diagonalizable. Otherwise, efficiently calculating
$\left[I-wE_{\rm s}D^{-1}A\right]^{-1}$ via matrix decomposition is difficult.
$D_{\rm s}^{-1}A_{\rm ss}$ is diagonalizable if the network is
undirected. A diagonalizable $D_{\rm s}^{-1}A_{\rm ss}$ possesses
$N_{\rm s}$ nondegenerate left
eigenvector $\bm v_{\rm s}^{(i)}$ ($1\le i\le N_{\rm s}$) with the corresponding eigenvalue
$\lambda_i$. It is possible that $\lambda_i=\lambda_j$ for $i\neq j$.

If $\lambda_i\neq 0$, $\lambda_i$ is an eigenvalue of
$E_{\rm s}D^{-1}A$, and the corresponding left eigenvector is given by
$\bm v^{(i)}=(\bm v_{\rm s}^{(i)}\; \bm v_{\rm d}^{(i)})$, where
\begin{equation}
\bm v_{\rm d}^{(i)} = \frac{\bm v_{\rm s}^{(i)}D_{\rm s}^{-1}A_{\rm sd}}{\lambda_i}.
\end{equation}

If $\lambda_i=0$, \EQ\eqref{eq:try left} implies that
$\bm v^{(i)}=(\bm v_{\rm s}^{(i)}\; \bm v_{\rm d}^{(i)})$ is not a
left eigenvector
of $E_{\rm s}D^{-1}A$. An example network with $N=3$ that has nontrivial zero eigenvalues
is presented in the next section for a pedagogical purpose.
When $\lambda_i=0$, I set $\bm v_{\rm
  d}^{(i)}=0$ such that
\begin{equation}
\bm v^{(i)}E_{\rm s}D^{-1}A =
(
\underbrace{\bm 0}_{N_{\rm s} \text{ zeros}}\;
\underbrace{\bm v_{\rm s}^{(i)}D_{\rm d}^{-1}A_{\rm sd}}_{\text{of size } N_{\rm d}}
).
\label{eq:nontrivial zero mode}
\end{equation}
Because $(\bm 0\quad \bm v_{\rm s}^{(i)}D_{\rm d}^{-1}A_{\rm sd})$ can be represented as a linear sum of
$\bm e_i^{\top}$ ($N_{\rm s}+1\le i\le N$), $\bm v^{(i)}$ is a type of generalized eigenvector
corresponding to $\lambda_i=0$.

I denote by $\bm u^{(i)}$ ($1\le i\le N_{\rm s}$) the nontrivial
generalized right eigenmodes
of $E_{\rm s}D^{-1}A$ corresponding to $\bm v^{(i)}$.
To obtain $\bm u^{(i)}$,
I denote by
$\bm u_{\rm s}^{(i)}$ ($1\le i\le N_{\rm s}$)
the normalized
right eigenvectors of $D_{\rm s}^{-1}A_{\rm ss}$ with eigenvalue $\lambda_i$.
Then,
\begin{equation}
\bm u^{(i)}\equiv
\begin{pmatrix}
\bm u_{\rm s}^{(i)}\\[4mm] \bm 0
\end{pmatrix}
\begin{array}{l}
\} \text{size } N_{\rm s}\\[4mm]
\} N_{\rm d} \text{ zeros}
\end{array}\quad (1\le i\le N_{\rm s})
\end{equation}
are right eigenvectors of
$E_{\rm s}D^{-1}A$ that respect the orthogonality
$\bm v^{(i)}\bm u^{(j)}=\delta_{ij}$, where $\delta$ is the Kronecker delta.

For completeness, I obtain the expression of the other $N_{\rm d}$ right eigenvectors
of $E_{\rm s}D^{-1}A$ corresponding to the trivial zero eigenvalue as follows.
I align
$\bm v^{(i)}$ and $\bm u^{(i)}$ ($1\le i\le N_{\rm s}$) such that nonzero eigenvectors
correspond to $1\le i\le N_{\rm s}-N_0$ and generalized zero eigenmodes correspond
to $N_{\rm s}-N_0+1\le i\le N_{\rm s}$. Then, the orthogonality condition
$\bm v^{(i)}\bm u^{(j)}=\delta_{ij}$ reads
\begin{equation}
\begin{pmatrix}
\bm v_{\rm s}^{(1)} & \frac{\bm v_{\rm s}^{(1)}D_{\rm s}^{-1}A_{\rm sd}}{\lambda_1}\\
\vdots & \vdots\\
\bm v_{\rm s}^{(N_{\rm s}-N_0)} & \frac{\bm v_{\rm s}^{(N_{\rm s}-N_0)}D_{\rm s}^{-1}A_{\rm sd}}{\lambda_{N_{\rm s}-N_0}}\\[4mm]
\bm v_{\rm s}^{(N_{\rm s}-N_0+1)} \\
\vdots & \bm 0\\
\bm v_{\rm s}^{(N_{\rm s})}\\[2mm]
\bm 0 & I_{N_{\rm d}}
\end{pmatrix}
\begin{pmatrix}
\bm u_{\rm s}^{(1)} \bm u_{\rm s}^{(2)} \cdots \bm u_{\rm s}^{(N_{\rm s})}
& M\\
0& I_{N_{\rm d}}
\end{pmatrix}
=I
\label{eq:M1}
\end{equation}
for an $N_{\rm s}$-by-$N_{\rm d}$ matrix $M$.
Equation~\eqref{eq:M1} yields
\begin{equation}
M = -
\begin{pmatrix}
\bm u_{\rm s}^{(1)} & \cdots & \bm u_{\rm s}^{(N_{\rm s}-N_0)}
\end{pmatrix}
\begin{pmatrix}
\frac{\bm v_{\rm s}^{(1)}D_{\rm s}^{-1}A_{\rm sd}}{\lambda_1}\\[2mm]
\vdots\\[2mm]
\frac{\bm v_{\rm s}^{(N_{\rm s}-N_0)}D_{\rm s}^{-1}A_{\rm sd}}{\lambda_{N_{\rm s}-N_0}}\end{pmatrix}.
\end{equation}

Finally, the decomposition of $E_{\rm s}D^{-1}A$ is given by
\begin{align}
&E_{\rm s}D^{-1}A\notag\\
=&
\begin{pmatrix}
\bm u_{\rm s}^{(1)} \bm u_{\rm s}^{(2)} \cdots \bm u_{\rm s}^{(N_{\rm s})}
& M\\[2mm]
\bm 0& I_{N_{\rm d}}
\end{pmatrix}
\begin{pmatrix}
\lambda_1 \bm v_{\rm s}^{(1)} & \bm v_{\rm s}^{(1)}D_{\rm s}^{-1}A_{\rm sd}\\
\vdots & \vdots\\
\lambda_{N_{\rm s}-N_0}\bm v_{\rm s}^{(N_{\rm s}-N_0)} & \bm v_{\rm s}^{(N_{\rm s}-N_0)}D_{\rm s}^{-1}A_{\rm sd}\\[4mm]
 & \bm v_{\rm s}^{(N_{\rm s}-N_0+1)}D_{\rm s}^{-1}A_{\rm sd}\\
\bm 0 & \vdots\\
 & \bm v_{\rm s}^{(N_{\rm s})}D_{\rm s}^{-1}A_{\rm sd}\\[2mm]
\bm 0 & \bm 0
\end{pmatrix}
\notag\\
=& \sum_{j=1}^{N_{\rm s}-N_0}\lambda_j
\begin{pmatrix} \bm u_{\rm s}^{(j)}\\[2mm] \bm 0\end{pmatrix}
\begin{pmatrix} \bm v_{\rm s}^{(j)}\quad \frac{\bm v_{\rm s}^{(j)}D_{\rm s}^{-1}A_{\rm sd}}
{\lambda_j}\end{pmatrix}
+
\sum_{j=N_{\rm s}-N_0+1}^{N_{\rm s}}
\begin{pmatrix} \bm u_{\rm s}^{(j)}\\[2mm] \bm 0\end{pmatrix}
\begin{pmatrix} \bm 0\quad \bm v_{\rm s}^{(j)}D_{\rm s}^{-1}A_{\rm sd}\end{pmatrix}.
\label{eq:expand 1}
\end{align}
Combining \EQ\eqref{eq:expand 1} and the orthogonality condition
$\bm v_{\rm s}^{(i)}\bm u_{\rm s}^{(j)}=\delta_{ij}$, 
I obtain
\begin{equation}
\left[E_{\rm s}D^{-1}A\right]^{\ell} =
\sum_{j=1}^{N_{\rm s}-N_0}\lambda_j^{\ell}
\begin{pmatrix} \bm u_{\rm s}^{(j)}\\[2mm] \bm 0\end{pmatrix}
\begin{pmatrix} \bm v_{\rm s}^{(j)}\quad \frac{\bm v_{\rm s}^{(j)}D_{\rm s}^{-1}A_{\rm sd}}
{\lambda_j}\end{pmatrix}\quad (\ell\ge 2).
\label{eq:expansion of ^l}
\end{equation}

Using \EQS\eqref{eq:E_s}, \eqref{eq:expand 1}, and \eqref{eq:expansion of ^l},
we can express the quantities appearing on the left-hand sides of
\EQS\eqref{eq:GC stable original} and \eqref{eq:stability cnd for all-GC 2}
as
\begin{align}
\left[I-w(I-E_i)D^{-1}A\right]^{-1} =& 
I+ \sum_{j=1}^{N_{\rm s}-N_0}\frac{w\lambda_j}{1-w\lambda_j}
\begin{pmatrix} \bm u_{\rm s}^{(j)}\\[2mm] \bm 0\end{pmatrix}
\begin{pmatrix} \bm v_{\rm s}^{(j)}\quad \frac{\bm v_{\rm s}^{(j)}D_{\rm s}^{-1}A_{\rm sd}}
{\lambda_j}\end{pmatrix}\notag\\
& + w
\sum_{j=N_{\rm s}-N_0+1}^{N_{\rm s}}
\begin{pmatrix} \bm u_{\rm s}^{(j)}\\[2mm] \bm 0\end{pmatrix}
\begin{pmatrix} \bm 0\quad \bm v_{\rm s}^{(j)}D_{\rm s}^{-1}A_{\rm sd}\end{pmatrix},\\
(I-E_i)D^{-1}A \left[I- w(I-E_i)D^{-1}A\right]^{-1} =&
\sum_{j=1}^{N_{\rm s}-N_0}\frac{\lambda_j}{1-w\lambda_j}
\begin{pmatrix} \bm u_{\rm s}^{(j)}\\[2mm] \bm 0\end{pmatrix}
\begin{pmatrix} \bm v_{\rm s}^{(j)}\quad \frac{\bm v_{\rm s}^{(j)}D_{\rm s}^{-1}A_{\rm sd}}
{\lambda_j}\end{pmatrix}\notag\\
& + 
\sum_{j=N_{\rm s}-N_0+1}^{N_{\rm s}}
\begin{pmatrix} \bm u_{\rm s}^{(j)}\\[2mm] \bm 0\end{pmatrix}
\begin{pmatrix} \bm 0\quad \bm v_{\rm s}^{(j)}D_{\rm s}^{-1}A_{\rm sd}\end{pmatrix}.
\end{align}

If $A_{\rm ss}$ is symmetric, $D_{\rm s}^{-1/2}A_{\rm ss}D_{\rm s}^{-1/2}$ is also symmetric and therefore
diagonalizable by a unitary matrix.
Denote the eigenvalue and the right eigenvector of
$D_{\rm s}^{-1/2}A_{\rm ss}D_{\rm s}^{-1/2}$ by $\hat{\lambda}_i$
and $\hat{\bm u}^{(i)}$,
respectively. Note that $\hat{\lambda}_i$ and $\hat{\bm u}^{(i)}$
are both real and can be computed
relatively easily.
Then, we can obtain the relationships
$\lambda_i=\hat{\lambda}_i$, $\bm u_{\rm s}^{(i)}=D_{\rm s}^{-1/2}\hat{\bm u}^{(i)}$, and
$\bm v_{\rm s}^{(i)}=
\hat{\bm u}^{(i)\top}D_{\rm s}^{1/2}$. We can 
also obtain $\bm v_{\rm d}=\hat{\bm u}^{(i)\top}D_{\rm s}^{-1/2}
A_{\rm sr}/\lambda_i$ when $\lambda_i\neq 0$.

\subsection*{Example network yielding nontrivial zero eigenmodes}

\begin{figure}
\begin{center}
\includegraphics[height=1.5cm]{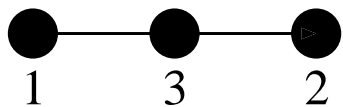}
\end{center}
\caption{{\bf A network yielding nontrivial zero eigenvalues.}}
\label{fig:example-N3-br}
\end{figure}

Consider the undirected network having $N=3$ nodes as shown in \FIG\ref{fig:example-N3-br}.
For this network I obtain
\begin{equation}
D^{-1}A = \begin{pmatrix}
0& 0& 1\\ 0& 0& 1\\ \frac{1}{2}& \frac{1}{2}& 0
\end{pmatrix}.
\end{equation}
By turning player 3 from GC to CD, I obtain
\begin{equation}
(I-E_3)D^{-1}A = (E_1+E_2)D^{-1}A = \begin{pmatrix}
0& 0& 1\\ 0& 0& 1\\ 0& 0& 0
\end{pmatrix}.
\label{eq:player 3 GC->CD}
\end{equation}
All of the eigenvalues of 
matrix~\eqref{eq:player 3 GC->CD}
are equal to zero, one trivial and two nontrivial.
The one trivial zero eigenvalue
originates from removing player 3 from the network of GCs.
The trivial zero left eigenvector is given by
$\bm v_3=\bm e_3^{\top}=(0\; 0\; 1)$.
%
%
I select the two generalized zero left eigenmodes to be
$\bm v_i=\bm e_i^{\top}$ ($i=1, 2$). The choice of $\bm v_1$
and $\bm v_2$ is not unique.
The right eigenmodes are given by $\bm u_i=\bm e_i (1\le i\le 3)$.

Equation~\eqref{eq:nontrivial zero mode}, for example, then reads
$\bm v_1 (E_1+E_2)D^{-1}A = \bm v_2 (E_1+E_2)D^{-1}A = \bm v_3$ and
$\bm v_3 (E_1+E_2)D^{-1}A=0$.

\section*{Acknowledgments}

I thank Hisashi Ohtsuki and Kazuo Murota for the helpful discussions and
acknowledge the support provided by
Grants-in-Aid for Scientific Research
(Grant Nos. 20760258 and 23681033, and
Innovative Areas ``Systems Molecular Ethology'') 
from MEXT, Japan.


\begin{thebibliography}{10}
\providecommand{\url}[1]{\texttt{#1}}
\providecommand{\urlprefix}{URL }
\expandafter\ifx\csname urlstyle\endcsname\relax
  \providecommand{\doi}[1]{doi:\discretionary{}{}{}#1}\else
  \providecommand{\doi}{doi:\discretionary{}{}{}\begingroup
  \urlstyle{rm}\Url}\fi
\providecommand{\bibAnnoteFile}[1]{%
  \IfFileExists{#1}{\begin{quotation}\noindent\textsc{Key:} #1\\
  \textsc{Annotation:}\ \input{#1}\end{quotation}}{}}
\providecommand{\bibAnnote}[2]{%
  \begin{quotation}\noindent\textsc{Key:} #1\\
  \textsc{Annotation:}\ #2\end{quotation}}
\providecommand{\eprint}[2][]{\url{#2}}

\bibitem{Nowak2005nature}
Nowak MA, Sigmund K (2005) Evolution of indirect reciprocity.
\newblock Nature 437: 1291--1298.
\input{Nowak2005nature}

\bibitem{Sigmund2010book}
Sigmund K (2010) The Calculus of Selfishness.
\newblock Princeton, NJ: Princeton University Press.
\input{Sigmund2010book}

\bibitem{Dufwenberg2001HomoOeco}
Dufwenberg M, Gneezy U, G{\"u}th W, van Damme E (2001) Direct vs indirect
  reciprocity: an experiment.
\newblock Homo Oecono 18: 19--30.
\input{Dufwenberg2001HomoOeco}

\bibitem{Greiner2005JEconPsy}
Greiner B, Levati MV (2005) Indirect reciprocity in cyclical networks---an
  experimental study.
\newblock J Econ Psych 26: 711--731.
\input{Greiner2005JEconPsy}

\bibitem{Stanca2009JEconPsy}
Stanca L (2009) Measuring indirect reciprocity: whose back do we scratch?
\newblock J Econ Psych 30: 190--202.
\input{Stanca2009JEconPsy}

\bibitem{Bartlett2006PsycSci}
Bartlett MY, DeSteno D (2006) Gratitude and prosocial behavior.
\newblock Psych Sci 17: 319--325.
\input{Bartlett2006PsycSci}

\bibitem{Rutte2007PlosBiol}
Rutte C, Taborsky M (2007) Generalized reciprocity in rats.
\newblock Plos Biol 5: e196.
\input{Rutte2007PlosBiol}

\bibitem{Yamagishi1993SPQ}
Yamagishi T, Cook KS (1993) Generalized exchange and social dilemmas.
\newblock Social Psychology Quarterly 56: 235--248.
\input{Yamagishi1993SPQ}

\bibitem{Molm2007AJS}
Molm LD, Collett JL, Schaefer DR (2007) Building solidarity through generalized
  exchange: a theory of reciprocity.
\newblock Am J Sociol 113: 205--242.
\input{Molm2007AJS}

\bibitem{Malinowski1922book}
Malinowski B (1922) Argonauts of the Western Pacific.
\newblock New York: E. P. Dutton.
\input{Malinowski1922book}

\bibitem{Ziegler1990chapter}
Ziegler R (1990) The kula: social order, barter, and ceremonial exchange.
\newblock In: Hechter M, Opp KD, Wippler R, editors, Social Institutions: Their
  Emergence, Maintenance, and Effects. New York: Aldine de Gruyter, pp.
  141--170.
\input{Ziegler1990chapter}

\bibitem{Barta2011RoyalB}
Barta Z, McNamara JM, Husz\'{a}r DB, Taborsky M (2011) Cooperation among
  non-relatives evolves by state-dependent generalized reciprocity.
\newblock Proc R Soc B 278: 843--848.
\input{Barta2011RoyalB}

\bibitem{Boyd1989SocN}
Boyd R, Richerson PJ (1989) The evolution of indirect reciprocity.
\newblock Soc Netw 11: 213--236.
\input{Boyd1989SocN}

\bibitem{Pfeiffer2005RoyalB}
Pfeiffer T, Rutte C, Killingback T, Taborsky M, Bonhoeffer S (2005) Evolution
  of cooperation by generalized reciprocity.
\newblock Proc R Soc B 272: 1115--1120.
\input{Pfeiffer2005RoyalB}

\bibitem{Nowak2007RoyalB}
Nowak MA, Roch S (2007) Upstream reciprocity and the evolution of gratitude.
\newblock Proc R Soc Lond B 274: 605--610.
\input{Nowak2007RoyalB}

\bibitem{Hamilton2005RoyalB}
Hamilton IM, Taborsky M (2005) Contingent movement and cooperation evolve under
  generalized reciprocity.
\newblock Proc R Soc B 272: 2259--2267.
\input{Hamilton2005RoyalB}

\bibitem{Rankin2009Evol}
Rankin DJ, Taborsky M (2009) Assortment and the evolution of generalized
  reciprocity.
\newblock Evolution 63: 1913--1922.
\input{Rankin2009Evol}

\bibitem{Iwagami2010JTB}
Iwagami A, Masuda N (2010) Upstream reciprocity in heterogeneous networks.
\newblock J Theor Biol 265: 297--305.
\input{Iwagami2010JTB}

\bibitem{Wasserman94}
Wasserman S, Faust K (1994) Social network analysis.
\newblock New York: Cambridge University Press.
\input{Wasserman94}

\bibitem{albert02rmp}
Albert R, Barab\'{a}si AL (2002) Statistical mechanics of complex networks.
\newblock Rev Mod Phys 74: 47--97.
\input{albert02rmp}

\bibitem{Newman03siam}
Newman MEJ (2003) The structure and function of complex networks.
\newblock SIAM Rev 45: 167--256.
\input{Newman03siam}

\bibitem{Newman10book}
Newman MEJ (2010) Networks --- An introduction.
\newblock Oxford: Oxford University Press.
\input{Newman10book}

\bibitem{Nowak1992Nature_spatial}
Nowak MA, May RM (1992) Evolutionary games and spatial chaos.
\newblock Nature 359: 826--829.
\input{Nowak1992Nature_spatial}

\bibitem{Santos2005PRL}
Santos FC, Pacheco JM (2005) Scale-free networks provide a unifying framework
  for the emergence of cooperation.
\newblock Phys Rev Lett 95: 098104.
\input{Santos2005PRL}

\bibitem{Santos2006PNAS}
Santos FC, Pacheco JM, Lenaerts T (2006) Evolutionary dynamics of social
  dilemmas in structured heterogeneous populations.
\newblock Proc Natl Acad Sci USA 103: 3490--3494.
\input{Santos2006PNAS}

\bibitem{Santos2006JEB}
Santos FC, Pacheco JM (2006) A new route to the evolution of cooperation.
\newblock J Evol Biol 19: 726--733.
\input{Santos2006JEB}

\bibitem{Duran2005PhysicaD}
Dur\'{a}n O, Mulet R (2005) Evolutionary prisoner's dilemma in random graphs.
\newblock Physica D 208: 257--265.
\input{Duran2005PhysicaD}

\bibitem{Tomassini2007IJMPC}
Tomassini M, Pestelacci E, Luthi L (2007) Social dilemmas and cooperation in
  complex networks.
\newblock Int J Mod Phys C 18: 1173--1185.
\input{Tomassini2007IJMPC}

\bibitem{Masuda2007RoyalB}
Masuda N (2007) Participation costs dismiss the advantage of heterogeneous
  networks in evolution of cooperation.
\newblock Proc R Soc Lond B 274: 1815--1821.
\input{Masuda2007RoyalB}

\bibitem{Trivers1971}
Trivers RL (1971) The evolution of reciprocal altruism.
\newblock Q Rev Biol 46: 35--57.
\input{Trivers1971}

\bibitem{Axelrod1984book}
Axelrod R (1984) Evolution of Cooperation.
\newblock NY: Basic Books.
\input{Axelrod1984book}

\bibitem{Watts1998Nature}
Watts DJ, Strogatz SH (1998) Collective dynamics of `small-world' networks.
\newblock Nature 393: 440--442.
\input{Watts1998Nature}

\bibitem{Barabasi99sci}
Barab\'{a}si AL, Albert R (1999) Emergence of scaling in random networks.
\newblock Science 286: 509--512.
\input{Barabasi99sci}

\bibitem{Klemm2002PRE1}
Klemm K, Egu\'{\i}luz VM (2002) Highly clustered scale-free networks.
\newblock Phys Rev E 65: 036123.
\input{Klemm2002PRE1}

\bibitem{Klemm2002PRE2}
Klemm K, Egu\'{\i}luz VM (2002) Growing scale-free networks with small-world
  behavior.
\newblock Phys Rev E 65: 057102.
\input{Klemm2002PRE2}

\bibitem{Motter2002pre}
Motter AE, Lai YC (2002) Cascade-based attacks on complex networks.
\newblock Phys Rev E 66: 065102.
\input{Motter2002pre}

\bibitem{Nowak2006Science}
Nowak MA (2006) Five rules for the evolution of cooperation.
\newblock Science 314: 1560--1563.
\input{Nowak2006Science}

\bibitem{Fortunato2006www}
Fortunato S, {Bogu\~{n}\'{a}} M, Flammini A, Menczer F (2006) How to make the
  top ten: approximating pagerank from in-degree.
\newblock Proc 4th Workshop on Algorithms and Models for the Web Graph (WAW
  2006) : 59--71.
\input{Fortunato2006www}

\bibitem{mkk09njp}
Masuda N, Kawamura Y, Kori H (2009) Impact of hierarchical modular structure on
  ranking of individual nodes in directed networks.
\newblock New J Phys 11: 113002.
\input{mkk09njp}

\bibitem{Ghoshal2011NatComm}
Ghoshal G, Barab\'{a}si AL (2011) Ranking stability and super-stable nodes in
  complex networks.
\newblock Nat Comm 2: 394.
\input{Ghoshal2011NatComm}

\end{thebibliography}

\clearpage


\clearpage

\clearpage

\clearpage

\clearpage

\end{document}